\documentclass[12pt, authoryear]{article}
\usepackage{amsfonts, graphicx,color,multirow, amsthm,amsmath,natbib}
\usepackage{setspace,url}
\usepackage{multirow}
\def\boxit#1{\vbox{\hrule\hbox{\vrule\kern6pt\vbox{\kern6pt#1\kern6pt}\kern6pt\vrule}\hrule}}

\begin{document}

\newtheorem{theorem}{\indent \sc Theorem}
\newtheorem{corollary}{\indent \sc Corollary}
\newtheorem{lemma}{\indent \sc Lemma}
\newtheorem{remark}{\indent \sc Remark}
\newtheorem{proposition}{\indent \sc Proposition}
\newcommand{\argmax}[0]{\mbox{argmax}}
\newcommand{\argmin}[0]{\mbox{argmin}}
\newcommand{\ol}[1]{\overline{#1}}

\newcommand{\var}{\mbox{var}}
\newcommand{\Var}{\mbox{Var}}
\newcommand{\bb}{\mbox{\bf b}}
\newcommand{\bff}{\mbox{\bf f}}
\newcommand{\bx}{\mbox{\bf x}}
\newcommand{\by}{\mbox{\bf y}}
\newcommand{\bg}{\mbox{\bf g}}
\newcommand{\bA}{\mbox{\bf A}}
\newcommand{\ba}{\mbox{\bf a}}
\newcommand{\bB}{\mbox{\bf B}}
\newcommand{\bC}{\mbox{\bf C}}
\newcommand{\bD}{\mbox{\bf D}}
\newcommand{\bE}{\mbox{\bf E}}
\newcommand{\bF}{\mbox{\bf F}}
\newcommand{\bG}{\mbox{\bf G}}
\newcommand{\bH}{\mbox{\bf H}}
\newcommand{\bI}{\mbox{\bf I}}
\newcommand{\bU}{\mbox{\bf U}}
\newcommand{\bV}{\mbox{\bf V}}
\newcommand{\bQ}{\mbox{\bf Q}}
\newcommand{\bR}{\mbox{\bf R}}
\newcommand{\bS}{\mbox{\bf S}}
\newcommand{\bX}{\mbox{\bf X}}
\newcommand{\bY}{\mbox{\bf Y}}
\newcommand{\bZ}{\mbox{\bf Z}}
\newcommand{\br}{\mbox{\bf r}}
\newcommand{\bv}{\mbox{\bf v}}
\newcommand{\bL}{\mbox{\bf L}}
\newcommand{\bbP}{\mathbb{P} }
\newcommand{\bone}{\mbox{\bf 1}}
\newcommand{\bsone}{\mbox{\scriptsize \bf 1}}
\newcommand{\bzero}{\mbox{\bf 0}}
\newcommand{\bveps}{\mbox{\boldmath $\varepsilon$}}
\newcommand{\bet}{\mbox{\boldmath $\eta$}}
\newcommand{\bxi}{\mbox{\boldmath $\xi$}}
\newcommand{\bzeta}{\mbox{\boldmath $\zeta$}}
\newcommand{\beps}{\mbox{\boldmath $\varepsilon$}}
\newcommand{\bbeta}{\mbox{\boldmath $\beta$}}
\newcommand{\balpha}{\mbox{\boldmath $\alpha$}}
\newcommand{\bPsi}{\mbox{\boldmath $\Psi$}}
\newcommand{\bomega}{\mbox{\boldmath $\omega$}}
\newcommand{\hbbeta}{\hat{\bbeta}}
\newcommand{\hbeta}{\hat{\beta}}
\newcommand{\Bbeta}{ \bar{\beta}}
\newcommand{\bmu}{\mbox{\boldmath $\mu$}}
\newcommand{\bgamma}{\mbox{\boldmath $\gamma$}}
\newcommand{\mv}{\mbox{V}}
\newcommand{\bSigma}{\mbox{\boldmath $\Sigma$}}
\newcommand{\cov}{\mbox{cov}}
\newcommand{\beq}{\begin{eqnarray}}
\newcommand{\eeq}{\end{eqnarray}}
\newcommand{\beqn}{\begin{eqnarray*}}
\newcommand{\eeqn}{\end{eqnarray*}}
\newcommand{\Mstar}{{\mathcal{M}_{\star}}}
\newcommand{\lammax}{{\lambda_{\max}}}
\newcommand{\A}{{\bbP_n \bPsi_jY}}
\newcommand{\B}{{(\bbP_n \bPsi_j\bPsi_j^T)^{-1}}}
\newcommand{\ea}{{E \bPsi_jY}}
\newcommand{\eb}{{(E \bPsi_j\bPsi_j^T)^{-1}}}

\title{\bf Nonparametric Independence Screening in Sparse Ultra-High Dimensional Additive Models
\thanks{Jianqing Fan is Frederick L. Moore Professor of Finance,
Department of Operations Research and Financial Engineering,
Princeton University, Princeton NJ 08544 (Email:
jqfan@princeton.edu). Yang Feng is Assistant Professor, Department of Statistics, Columbia University, New York, NY 10027 (Email: yangfeng@stat.columbia.edu). Rui Song is
Assistant Professor, Department of Statistics, Colorado State
University, Fort Collins, CO 80523 (Email: song@stat.colostate.edu).
The financial support from
NSF grants DMS-0714554, DMS-0704337, DMS-1007698 and NIH grant R01-GM072611 are
greatly acknowledged. The authors are in deep debt to Dr. Lukas
Meier for sharing the codes of penGAM. The authors thank the editor,
the associate editor, and referees for their constructive comments.}}

\author{Jianqing Fan, Yang Feng and Rui Song}
\maketitle


%
%
%
%
%
%
%

\begin{abstract}
\begin{singlespace}
A variable screening procedure via correlation learning was proposed in Fan and Lv (2008) to reduce
dimensionality in sparse ultra-high dimensional models. Even when the true model is linear, the
marginal regression can be highly nonlinear.  To address this issue, we further extend the
correlation learning to marginal nonparametric learning. Our nonparametric independence screening is called NIS, a specific member of the sure independence screening.  Several closely related variable screening
procedures are proposed. Under general nonparametric models, it
is shown that under some mild technical conditions, the proposed independence screening methods
enjoy a sure screening property. The extent to which the dimensionality can be reduced by
independence screening is also explicitly quantified. As a methodological extension, a data-driven thresholding and an iterative
nonparametric independence screening (INIS) are also proposed to enhance the finite sample performance for fitting sparse additive models.  The simulation results and a real data analysis
demonstrate that the proposed procedure works well with moderate sample size and large dimension and
performs better than competing methods.
\end{singlespace}
\end{abstract}



\section{Introduction}
With rapid advances of computing power and other modern technology, high-throughput data of
unprecedented size and complexity are frequently seen in many contemporary statistical studies.
Examples include data from genetic, microarrays, proteomics, fMRI, functional data and high
frequency financial data. In all these examples, the number of variables $p$ can grow much faster
than the number of observations $n$. To be more specific, we assume $\log p=O(n^a)$ for some $a\in
(0,1/2)$. Following \cite{FanLv09}, we call it non-polynomial (NP) dimensionality or ultra-high
dimensionality.  What makes the under-determined statistical inference possible is the sparsity
assumption: only a small set of independent variables contribute to the response. Therefore,
dimension reduction and feature selection play pivotal roles in these ultra-high dimensional
problems.

The statistical literature contains numerous procedures on the
variable selection for linear models and other parametric models,
such as the Lasso \citep{Tibs96}, the SCAD and other folded-concave
penalty \citep{Fan97,FanLi01}, the Dantzig selector \citep{CT07},
the Elastic net (Enet) penalty \citep{ZouHastie05}, the MCP
\citep{Zhang09} and related methods \citep{Zou06,ZouLi08}.
Nevertheless, due to the ``curse of dimensionality'' in terms of
simultaneous challenges on the computational expediency,
statistical accuracy and algorithmic stability, these methods meet
their limits in ultra-high dimensional problems.

Motivated by these concerns, \cite{FanLv08} introduced a new framework for variable screening via
correlation learning with NP-dimensionality in the context of least squares. \cite{HTX09} used a
different marginal utility, derived from an empirical likelihood point of view. \cite{HM09}
proposed a generalized correlation ranking, which allows nonlinear regression. \cite{Huang08} also
investigated the marginal bridge regression in the ordinary linear model. These methods focus on
studying the marginal pseudo-likelihood and are fast but crude in terms of reducing the
NP-dimensionality to a more moderate size. To enhance the performance, \cite{FanLv08} and
\cite{Fan08} introduced some methodological extensions including iterative SIS (ISIS) and multi-stage
procedures, such as SIS-SCAD and SIS-LASSO, to select variables and estimate parameters
simultaneously.  Nevertheless, these marginal screening methods have some methodological
challenges. When the covariates are not jointly normal, even if the linear model holds in the joint
regression, the marginal regression can be highly nonlinear.  Therefore, sure screening based on
nonparametric marginal regression becomes a natural candidate.

In practice, there is often little prior information that the
effects of the covariates take a linear form or belong to any other
finite-dimensional parametric family. Substantial improvements are
sometimes possible by using a more flexible class of nonparametric
models, such as the additive model $Y = \sum_{j=1}^p m_j(X_{j}) +
\varepsilon$, introduced by \cite{Stone85}. It increases
substantially the flexibility of the ordinary linear model and
allows a data-analytic transform of the covariates to enter into the
linear model. Yet, the literature on variable selection in
nonparametric additive models are limited. See, for example,
\cite{KY08}, \cite{Ravikumar09}, \cite{Huang09} and
\cite{MeierGeerBuhlmann09}. \cite{KY08} and \cite{Ravikumar09} are
closely related with COSSO proposed in \cite{LZ06} with fixed
minimal signals, which does not converge to zero. \cite{Huang09} can
be viewed as an extension of adaptive lasso to additive models with
fixed minimal signals. \cite{MeierGeerBuhlmann09} proposed a penalty
which is a combination of sparsity and smoothness with a fixed
design. Under ultra-high dimensional settings, all these methods
still suffer from the aforementioned three challenges as they can be
viewed as extensions of penalized pseudo-likelihood approaches to
additive modeling. The commonly used algorithm in additive modeling
such as backfitting makes the situation even more challenging, as it
is quite computationally expensive.

In this paper, we consider independence learning by ranking the
magnitude of marginal estimators, nonparametric marginal
correlations, and the marginal residual sum of squares. That is, we
fit $p$ marginal nonparametric regressions of the response $Y$
against each covariate $X_i$ separately and rank their importance to
the joint model according to a measure of the goodness of fit of
their marginal model. The magnitude of these marginal utilities can
preserve the non-sparsity of the joint additive models under some
reasonable conditions, even with converging minimum strength of
signals. Our work can be regarded as an important and nontrivial
extension of SIS procedures proposed in \cite{FanLv08} and
\cite{FanSong09}. Compared with these papers, the minimum
distinguishable signal is related with not only the stochastic error
in estimating the nonparametric components, but also approximation
errors in modeling nonparametric components, which depends on the
number of basis functions used for the approximation. This brings
significant challenges to the theoretical development and leads to
an interesting result on the extent to which the dimensionality can
be reduced by nonparametric independence screening. We also propose
an iterative nonparametric independence screening procedure,
INIS-penGAM, to reduce the false positive rate and stabilize the
computation. This two-stage procedure can deal with the
aforementioned three challenges better than other methods, as will
be demonstrated in our empirical studies.

We approximate the nonparametric additive components by using a B-spline basis. Hence, the
component selection in additive models can be viewed as a functional version of the grouped
variable selection. An early literature on the group variable selection using group penalized
least-squares is \cite{AntoFan01} (see page 966), in which blocks of wavelet coefficients are
either killed or selected. The group variable selection was more thoroughly studied
in \cite{YuanLin06}, \cite{Kim06}, \cite{WeiHuang07} and \cite{MeierGeerBuhlmann09}.  Our methods
and results have important implications on the group variable selections, as in additive regression, each component can be expressed as a linear combination of a set of basis functions, whose coefficients
have to be either killed or selected simultaneously.

The rest of the paper is organized as follows.  In Section 2, we introduce the nonparametric
independence screening (NIS) procedure in additive models. The theoretical properties for NIS are
presented in Section 3.  As a methodological extension, INIS-penGAM and its greedy version g-INIS-penGAM are outlined in Section 4.
Monte Carlo simulations and a real data analysis in Section 5 demonstrate the effectiveness of the
INIS method. We conclude with a discussion in Section 6 and relegate the proofs to Section 7.

\section{Nonparametric independence screening}

Suppose that we have a random sample  $\{(\bX_i, Y_i)\}_{i=1}^n$
from the population \beq \label{eq1}
   Y = m(\bX) + \varepsilon,
\eeq
in which $\bX=(X_1, \ldots, X_p)^T$, $\varepsilon$ is the random error with conditional
mean zero.  To expeditiously identify important variables in model
(\ref{eq1}), without the ``curse-of-dimensionality'', we consider
the following $p$ marginal nonparametric regression problems:
\beq\label{eq2} \min_{f_j \in L_2(P)} E\Bigl(Y - f_j(X_j)\Bigr)^2,
\eeq where $P$ denotes the joint distribution of $(\bX, Y)$ and
$L_2(P)$ is the class of square integrable functions
under the measure $P$. The minimizer of (\ref{eq2}) is $f_j =
E(Y|X_j)$, the projection of $Y$ onto $X_j$. We rank the utility of
covariates in model (\ref{eq1}) according to, for example,
$E f_j^2(X_j)$ and select a small group of covariates
via thresholding.

To obtain a sample version of the marginal nonparametric regression, we employ a B-Spline basis.
Let $\mathcal{S}_n$ be the space of polynomial splines of degree $l \ge 1$ and
$\{\Psi_{jk},~k=1,\cdots,d_n\}$ denote a normalized B-Spline basis with $\|\Psi_{jk}\|_{\infty} \le
1$, where $\|\cdot \|_{\infty}$ is the sup norm. For any $f_{nj} \in \mathcal{S}_n$, we have
 \beqn
   f_{nj}(x) = \sum_{k=1}^{d_n} \beta_{jk}\Psi_{jk}(x), ~1 \le j \le p,
 \eeqn
for some coefficients $\{\beta_{jk}\}_{k=1}^{d_n}$.  Under some
smoothness conditions, the nonparametric projections
$\{f_j\}_{j=1}^p$ can well be approximated by functions in
$\mathcal{S}_n$.  The sample version of the marginal regression
problem can be expressed as \beq\label{eq3}
  \min_{f_{nj} \in \mathcal{S}_n} \mathbb{P}_n\Bigl(Y
        -  f_{nj}(X_{j})\Bigr)^2= \min_{\bbeta_j \in \mathbb{R}^{d_n}}
  \mathbb{P}_n \Bigl ( Y -  \bPsi_j^T
\bbeta_j \Bigr )^2, \eeq where $\bPsi_j \equiv \bPsi_j(X_j)=(\Psi_1(X_j), \cdots,
\Psi_{d_n}(X_j))^T$ denotes the $d_n$ dimensional basis functions and $\mathbb{P}_n g(\bX, Y)$ is
the expectation with respect to the empirical measure $\mathbb{P}_n$, i.e., the sample average of
$\{g(\bX_i, Y_i)\}_{i=1}^n$. This univariate nonparametric smoothing can be rapidly computed, even for
NP-dimensional problems. We correspondingly define the population version of the minimizer of the
componentwise least square regression, \beqn f_{nj} (X_j) = \bPsi_j^T(E\bPsi_j \bPsi_j^T)^{-1} E \bPsi_jY,
\qquad j=1,\cdots, p. \eeqn where $E$ denotes the expectation under the true model.

We now select a set of variables
\beq\label{eq4} \widehat {\cal{M}}_{\nu_n}=\{1\le j \le p: \|\hat
f_{nj} \|_n^2 \ge \nu_n \}, \eeq where $\|\hat f_{nj}\|_n^2 =
n^{-1} \sum_{i=1}^n \hat f_{nj}(X_{ij})^2$ and $\nu_n$ is a
predefined threshold value. Such an independence screening ranks the
importance according to the marginal strength of the marginal
nonparametric regression. This screening can also be viewed as
ranking by the magnitude of the correlation of the marginal
nonparametric estimate $\{\hat f_{nj}(X_{ij})\}_{i=1}^n$ with the
response $\{Y_i\}_{i=1}^n$, since $\|\hat f_{nj} \|_n^2 = \|Y\hat
f_{nj} \|_n$.  In this sense, the proposed NIS procedure
is related to the correlation learning proposed in \cite{FanLv08}.  

Another screening approach is to rank according to the descent order
of the residual sum of squares of the componentwise nonparametric
regressions, where we select a set of variables: \beqn
    \widehat {\cal{N}}_{\gamma_n}=\{1\le j \le p: u_j \le \gamma_n \},
\eeqn with $u_j = \min_{\bbeta_j} \bbP_n(Y - \bPsi_j^T \bbeta_j)^2$ is the residual sum of squares of the marginal fit and $\gamma_n$ is a predefined threshold value. It is straightforward to
show that $u_j = \bbP_n (Y^2 - \hat f_{nj}^2)$. Hence, the two methods are equivalent.

The nonparametric independence screening reduces the dimensionality
from $p$ to a possibly much smaller space with model size $|\widehat
{\cal{M}}_{\nu_n}|$ or $|\widehat {\cal{N}}_{\gamma_n}|$. It is
applicable to all models.  The question is whether we have
mistakenly deleted some active variables in model (\ref{eq1}).  In
other words, whether the procedure has a sure screening property as
postulated by \cite{FanLv08}.  In the next section, we will show
that the sure screening property indeed holds for nonparametric
additive models with a limited false selection rate.

\section{Sure Screening Properties}
In this section, we establish the sure screening properties for additive models with results presented in three steps.

\subsection{Preliminaries}

We now assume that the true regression function admits the additive
structure: \beq \label{eq5} m(\bX) = \sum_{j=1}^p m_j(X_j). \eeq For
identifiability, we assume $\{m_j(X_j)\}_{j=1}^p$ have mean zero.
Consequently, the response $Y$ has zero mean, too. Let
$\mathcal{M}_{\star}=\{j: Em_j(X_j)^2>0 \}$ be the true sparse
model with non-sparsity size $s_n = |\cal{M}_{\star}|$. We allow $p$ to grow with $n$ and
denote it as $p_n$ whenever needed.

The theoretical basis of the sure screening is that the marginal
signal of the active components ($\|f_{j}\|, j \in
\mathcal{M}_{\star}$) does not vanish, where $\|f_{j}\|^2 =
Ef_{j}^2$.  The following conditions make this possible.  For
simplicity, let $[a, b]$ be the support of $X_j$.
\begin{itemize}
\item[A.] The nonparametric marginal projections $\{f_j\}_{j=1}^p$ belong to a class of functions $\cal{F}$ whose $r$th derivative $f^{(r)}$ exists and is Lipschitz of order $\alpha$:
\beqn
\mathcal{F}=\Bigl\{f(\cdot):~\Bigl|f^{(r)}(s) - f^{(r)}(t)\Bigr| \le K|s-t|^{\alpha},~\mbox{for}~s,t\in[a,b]\Bigr\},
\eeqn
for some positive constant $K$,
where $r$ is a non-negative integer and $\alpha\in(0,1]$ such that $d=r+\alpha>0.5$.
  \item [B.] The marginal density function $g_j$
  of $X_j$ satisfies $0 < K_1 \le g_j(X_j) \le K_2 < \infty$ on $[a,b]$ for $1\le j\le p$ for some constants $K_1$ and $K_2$.

\item [C.] $\min_{j \in \Mstar} E\{E(Y|X_j)^2\}\geq c_1d_n n^{-2\kappa}$, for some
  $0 < \kappa < d/(2d+1)$ and $c_1 > 0$.
\end{itemize}
Under conditions A and B, the following three
facts hold when $l \geq d$ and will be used in the paper. We state
them here for readability.
\begin{itemize}
\item[Fact~1.]  There exists a positive constant $C_1$ such that \citep{Stone85}
\beq\label{eq6} \|f_j-f_{nj}\|^2 \le C_1 d_n^{-2d}. \eeq
\item[Fact~2.]  There exists a positive constant $C_2$ such that \citep{Stone85,Huang09}
\beq\label{eq7} E\Psi_{jk}^2(X_{ij}) \le C_2 d_n^{-1}. \eeq
\item[Fact~3.] There exist some positive constants $D_{1}$ and $D_{2}$ such that \citep{Zhou98}
\beq\label{eq8} D_{1} d_n^{-1} \le \lambda_{\min} (E\bPsi_j
\bPsi_j^T ) \le \lambda_{\max} (E\bPsi_j \bPsi_j^T ) \le D_{2}
d_n^{-1}. \eeq
\end{itemize}

The following lemma shows that the minimum signal of  $\{\|f_{nj}
\|\}_{j\in \cal{M}_*}$ is at the same level of the marginal
projection, provided that the approximation error is negligible.

\begin{lemma}\label{lem-0}
Under conditions A--C, we have \beqn min_{j\in \Mstar} \|f_{nj} \|^2
\geq c_1\xi  d_n n^{-2\kappa}, \eeqn
provided that $ d_n^{-2d-1} \le
c_1(1-\xi)n^{-2\kappa}/C_1$ for some $\xi \in (0,1)$.
\end{lemma}

A model selection consistency result can be established with nonparametric independence screening
under the partial orthogonality condition, i.e.,  $\{X_j, ~j \notin \mathcal{M}_{\star}\}$ is
independent of $\{ X_i, ~i \in \mathcal{M}_{\star}\}$.  In this case, there is a separation between
the strength of marginal signals $\|f_{nj}\|^2$ for active variables $
\{X_j; j \in \mathcal{M}_{\star}\}$ and  inactive variables $\{X_j, j \notin \mathcal{M}_{\star}\}$, which are zero.  When the separation is sufficiently large, these two sets of variables can
be easily identified.


\subsection{Sure Screening}

In this section, we establish the sure screening properties of the nonparametric independence
screening (NIS). We need the following additional conditions:
\begin{itemize}
  \item [D.]   $\|m \|_{\infty} < B_1$ for some positive constant $B_1$, where $\|\cdot \|_{\infty}$ is the sup norm.
  \item [E.] The random error $\{\varepsilon_i\}_{i=1}^n$ are i.i.d. with conditional  mean zero
  and  for any $B_2>0$, there exists a positive constant $B_3$ such that $E[\exp(B_2 |\varepsilon_i|)|\bX_i] < B_3$.
  \item [F.] There exist a positive constant $c_1$ and $\xi \in (0,1)$ such that $d_n^{-2d-1} \le c_1(1-\xi)n^{-2\kappa}/C_1$.
\end{itemize}

The following theorem gives the sure screening properties. It reveals that
it is only the size of non-sparse elements $s_n$ that matters for the purpose of sure screening, not the dimensionality $p_n$.  The first result is on the uniform convergence of $\|\hat f_{nj}\|_n^2$ to $\|f_{nj}\|^2$.

\begin{theorem}\label{the-1}
Suppose that Conditions A, B, D and E hold.
\begin{itemize}
\item[(i)] For any $c_2>0$, there exist some positive constants $c_3$ and $c_4$ such that
\beq \label{eq9}
&&P\Bigl(\max_{1\le j\le p_n}\Bigl| \|\hat f_{nj}\|_n^2 - \|f_{nj}\|^2 \Bigr| \ge
c_2d_n n^{-2\kappa}\Bigr) \nonumber \\
&\le&
  p_n d_n \Bigl\{(8+2d_n)\exp\Bigl(-c_3n^{1-4\kappa}d_n^{-3}\Bigr) +
6d_n \exp\Bigl(-c_4nd_n^{-3}\Bigr)\Bigr\}.
\eeq
\item[(ii)] If, in addition, Conditions C and F hold, then by taking $\nu_n = c_5d_nn^{-2\kappa}$ with $c_5 \le c_1 \xi/2$, we have
\beqn
P(\mathcal{M}_{\star} \subset \widehat {\cal{M}}_{\nu_n}) &\ge&
1 -
 s_n d_n \Bigl\{(8+2d_n)\exp\Bigl(-c_3n^{1-4\kappa}d_n^{-3}
\Bigr) \\
&&+ 6d_n \exp\Bigl(-c_4nd_n^{-3}\Bigr)  \Bigr\}.
\eeqn
\end{itemize}
\end{theorem}

Note that the second part of the upper bound in Theorem \ref{the-1} is related to the uniform
convergence rates of the minimum eigenvalues of the design matrices. It gives an upper bound on the
number of basis $d_n=o(n^{1/3})$ in order to have the sure screening property, whereas Condition F requires $d_n \geq B_4 n^{2\kappa/(2d+1)}$, where $B_4= (c_1(1-\xi)/C_1)^{-1/(2d+1)}$.

It follows from Theorem \ref{the-1} that we can handle the
NP-dimensionality:
\begin{eqnarray} \label{eq10}
\log p_n = o(n^{1-4\kappa}d_n^{-3} + nd_n^{-3}).
\end{eqnarray}
Under this condition,
$$
       P(\mathcal{M}_{\star} \subset \widehat {\cal{M}}_{\nu_n}) \to 1,
$$
i.e., the sure screening property. It is worthwhile to point out
that the number of spline basis $d_n$ affects the order of
dimensionality, comparing with the results of \cite{FanLv08} and
\cite{FanSong09} in which univariate marginal regression is used.
Equation (\ref{eq10}) shows that the larger the minimum signal level
or the smaller the number of basis functions, the higher
dimensionality the nonparametric independence screening (NIS) can
handle. This is in line with our intuition. On the other hand, the
number of basis functions can not be too small, since the
approximation error can not be too large. As required by Condition
F, $d_n \ge B_4 n^{2\kappa/(2d+1)}$; the smoother the underlying
function, the smaller $d_n$ we can take and the higher the dimension
that the NIS can handle. If the minimum signal does not converge to
zero, as in \cite{LZ06}, \cite{KY08} and \cite{Huang09}, then
$\kappa = 0$. In this case, $d_n$ can be taken to be finite as long
as it is sufficiently large so that minimum signal in Lemma 1
exceeds the noise level. By
taking $d_n = n^{1/(2d+1)}$, the optimal rate for nonparametric
regression \citep{Stone85}, we have $\log p_n = o(
n^{2(d-1)/(2d+1)})$. In other words, the dimensionality can be as
high as $\exp\{o( n^{2(d-1)/(2d+1)})\}$.

\subsection{Controlling false selection rates}

The sure screening property, without controlling false selection
rates, is not insightful.  It basically states that the NIS has no
false negatives. An ideal case for the vanishing false positive rate
is that \beqn \max_{j \notin \mathcal{M}_{\star}}\|f_{nj}\|^2 =
o(d_n n^{-2\kappa}),\eeqn
so that there is a gap between active variables and inactive variables
in model (\ref{eq1}) when using the marginal nonparametric screener.
In this case, by Theorem \ref{the-1}(i), if (\ref{eq9}) tends to
zero, with probability tending to one that \beqn \max_{j \notin
\mathcal{M}_{\star}} \|\hat f_{nj}\|^2_n \leq c_2 d_n n^{-2\kappa},
\qquad \mbox{for any $c_2 > 0$.} \eeqn
Hence, by the
choice of $\nu_n$ as in Theorem~\ref{the-1}(ii), we can achieve
model selection consistency:
$$
      P( \widehat
{\mathcal{M}}_{\nu_n} = \mathcal{M}_{\star} ) = 1 - o(1).
$$

We now deal with the more general case.  The idea is to bound the size of the selected set by using
the fact that $\var(Y)$ is bounded. In this part, we show that the correlations among the basis
functions, i.e., the design matrix of the basis functions, are related to the size of selected models.


\begin{theorem}\label{the-2}
Suppose Conditions A--F hold and $\var(Y)=O(1)$.
Then, for any $\nu_n = c_5 d_nn^{-2\kappa}$, there exist positive
constants $c_3$ and $c_4$ such that \beqn && P [ | \widehat {\cal
M}_{\nu_n}|  \leq
O\{ n^{2\kappa}\lambda_{\max} (\bSigma)\} ]\\
& \geq&  1 - p_n d_n \Bigl\{(8+2d_n)\exp(-c_{3}n^{1-4\kappa}d_n^{-3}) +
6d_n\exp(-c_4nd_n^{-3})\Bigr\}, \eeqn where $\bSigma =
E\bPsi\bPsi^T$ and $\bPsi = (\bPsi_1, \cdots, \bPsi_{p_n})^T.$
\end{theorem}

The significance of the result is that when
$\lambda_{\max}(\bSigma)=O(n^{\tau})$, the selected model size with
the sure screening property is only of polynomial order, whereas the
original model size is of NP-dimensionality.  In other words, the
false selection rate converges to zero exponentially fast.
The size of the selected
variables is of order $O(n^{2\kappa+\tau})$. This is of the same
order as in Fan and Lv (2008). Our result is an extension of
\cite{FanLv08}, even in this very specific case without the
condition $2\kappa + \tau < 1$.
The results are also consistent with that in
\cite{FanSong09}: the number of selected variables is related to the
correlation structure of the covariance matrix.

In the specific case where the covariates are independent, then the
matrix $\bSigma$ is block diagonal with $j$-th block $\bSigma_j$.
Hence, it follows from (\ref{eq8}) that $\lambda_{\max}(\bSigma) =
O(d_n^{-1})$.

\section{INIS Method}
\subsection{Description of the Algorithm}
After variable screening, the next step is naturally to select the
variables using more refined techniques in the additive model. For
example, the penalized method for additive model (penGAM) in
\cite{MeierGeerBuhlmann09} can be employed to select a subset of
active variables.  This results in NIS-penGAM.  To further enhance
the performance of the method, in terms of false selection rates,
following \cite{FanLv08} and \cite{Fan08}, we can iteratively employ
the large-scale screening and moderate-scale selection
strategy, resulting in the INIS-penGAM.  

Given the data $\{(\bX_i,Y_i)\}, i=1,\cdots,n$, for each component $f_j(\cdot), j=1,\cdots, p$, we
choose the same truncation term $d_n=O(n^{1/5})$. To determine a data-driven thresholding for independence screening,  we extend the random permutation idea in \citet{ZhaoLi10}, which allows only $1-q$ proportion (for a given $q\in [0,1]$) of inactive variables to enter the model when $\bX$ and $Y$ are not related (the null model).  The random permutation is used to decouple $\bX_i$ and $Y_i$ so that the resulting data $(\bX_{\pi(i)}, Y_i)$ follow a null model, where $\pi(1), \cdots, \pi(n)$ are a random permutation of the index $1, \cdots, n$. The algorithm works as follows:

\begin{enumerate}
\item[Step 1:] For every $j\in \{1,\cdots, p\}$, we compute
\beqn
  \hat f_{nj} =\argmin_{f_{nj} \in \mathcal{S}_n} \mathbb{P}_n\Bigl(Y -
    f_{nj}(X_{j})\Bigr)^2, ~ \mbox{for}~ 1 \le j \le p.
 \eeqn
Randomly permute the rows of $\bX$, yielding $\tilde\bX$.
Let $\omega_{(q)}$ be the $q^{th}$ quantile of $\{\|\hat f^*_{nj}\|_n^2, j=1,2,\cdots,p\}$, where
\beqn
  \hat f^*_{nj} =\argmin_{f_{nj} \in \mathcal{S}_n} \mathbb{P}_n\Bigl(Y -
    f_{nj}(\tilde X_{j})\Bigr)^2.
 \eeqn
Then, NIS selects the following variables:
$$
  \mathcal{A}_1 =\{j:  \|\hat f^*_{nj}\|_n^2\geq \omega_{(q)}\}.
$$
In our numerical examples, we use $q=1$ (i.e., take the maximum value of the empirical norm of the permuted estimates).

\item[Step 2:] We apply further the penalized method for additive model (penGAM) in \cite{MeierGeerBuhlmann09}
on the set $\mathcal{A}_1$ to select a subset $\mathcal{M}_1$. Inside the penGAM algorithm, the
penalty parameter is selected by cross validation.
\item[Step 3:] For every $j\in \mathcal{M}_1^c=\{1,\cdots, p\} \backslash
\mathcal{M}_1$, we minimize \beq \label{eq11}
   \mathbb{P}_n\Bigl(Y - \sum_{i\in \mathcal{M}_1}f_{ni}(X_i)-
        f_{nj}(X_{j})\Bigr)^2,
\eeq
with respect to $f_{ni} \in \mathcal{S}_n$ for all $i \in \mathcal{M}_1$ and $f_{nj} \in \mathcal{S}_n$.
This regression reflects the additional contribution of the $j$-th
components conditioning on the existence of the variable set
$\mathcal{M}_1$. After marginally screening as in the first step, we
can pick a set $\mathcal{A}_2$ of indices. Here the size determination is the same as in Step 1, except that only the variables not in $\mathcal{M}_1$ are randomly permuted. Then we apply further the penGAM
algorithm on the set $\mathcal{M}_1 \bigcup \mathcal{A}_2$ to select
a subset $\mathcal{M}_2$.
\item[Step 4:] We iterate the process until $|\mathcal{M}_l|\geq s_0$ or $\mathcal{M}_{l}=\mathcal{M}_{l-1}$.
\end{enumerate}

Here are a few comments about the method.  In Step 2, we use the penGAM method. In fact, any
variable selection method for additive models will work such as the SpAM in \citet{Ravikumar09} and
also the adaptive group LASSO for additive models in \citet{Huang09}. A similar sample splitting
idea as described in \citet{Fan08} can be applied here to further reduce false selection rate.
\subsection{Greedy INIS (g-INIS)}
We now propose a greedy modification to the INIS algorithm to speed up the computation and to enhance the performance. Specifically, we restrict the size of the set ${\cal A}_j$ in the iterative screening steps to be at most $p_0$, a small positive integer, and the algorithm stops when none of the variables is recruited, i.e., exceeding the thresholding for the null model.  In the numerical studies, $p_0$ is taken to be one for simplicity.   This greedy version of the INIS algorithm is called ``g-INIS".

When $p_0=1$, the g-INIS method is connected with the forward selection \citep{Efroymson-1960, DraperSmith-1966}. Recently, \citet{WANG2009} showed that under certain technical conditions, forward selection can also achieve the sure screening property. Both g-INIS and forward selection recruit at most one new variable into the model at a time. The major difference is that unlike the forward selection which keeps a variable once selected, g-INIS has a deletion step via penalized least-squares that can remove multiple variables. This makes the g-INIS algorithm more attractive since it is more flexible in terms of recruiting and deleting variables.

The g-INIS is particularly effective when the covariates are highly correlated or conditionally correlated. In this case,  the original INIS method tends to select many  unimportant variables that have high correlation with important variables as they, too, have large marginal effects on the response.  Although greedy, the g-INIS method is better at choosing true positives due to more stringent screening and improves the chance of the remaining important variables to be selected in subsequent stages due to less false positives at each stage.  This leads to conditioning on a smaller set of more relevant variables and improve the overall performance.  From our numerical experience, the g-INIS method outperforms the original INIS method in all examples in terms of higher true positive rate, smaller false positive rate and smaller prediction error.

\section{Numerical Results}
In this section, we will illustrate our method by studying the
performance on the simulated data and a real data analysis. Part of
the simulation settings are adapted from \cite{FanLv08},
\cite{MeierGeerBuhlmann09}, \cite{Huang09}, and \cite{FanSong09}.

\subsection{Comparison of Minimum Model Size}
We first illustrate the behavior of the NIS procedure under
different correlation structures.  Following \cite{FanSong09}, the
minimum model size(MMS) required for the NIS procedure and the
penGAM procedure to have the sure screening property, i.e., to
contain the true model $\mathcal{M}^*$, is used as a measure of the
effectiveness of a screening method. We also include the correlation
screening of \cite{FanLv08} for comparison. The advantage of the MMS
method is that we do not need to choose the thresholding parameter
or penalized parameters.
 For NIS, we take $d_n = \lfloor n^{1/5}\rfloor+2=5$.  We set $n=400$ and $p=1000$ for all examples.

{\bf Example 1}.  Following \cite{FanSong09}, let $\{X_k\}_{k=1}^{950}$ be i.i.d standard normal random variables and
$$
X_k=\sum_{j=1}^sX_j(-1)^{j+1}/5+\sqrt{1-\frac{s}{25}}\varepsilon_k,
\quad\quad k=951, \cdots, 1000,
$$
where $\{\varepsilon_k\}_{k=951}^{1000}$ are standard normally
distributed. We consider the following linear model as a specific case of the additive model: $Y={\bbeta^*}^{T}\bX+\varepsilon$, in which $\varepsilon\sim N(0,3)$ and $\bbeta^*=(1,-1,\cdots)^T$ has $s$ non-vanishing components, taking values $\pm 1$ alternately.


{\bf Example 2}.  In this example, the data is generated from the simple linear regression $Y=X_{1}+X_{2}+X_{3}+\varepsilon$, where $\varepsilon\sim N(0,3)$.  However, the covariates are not normally distributed:
$\{X_{k}\}_{k\neq 2}$ are i.i.d standard normal random variables whereas $X_{2}=-\frac{1}{3}X_{1}^3+\tilde\varepsilon$, where $\tilde\varepsilon\sim N(0,1)$.  In this case, $E(Y|X_1)$ and $E(Y|X_2)$ are nonlinear.

\begin{table}[ht]
\caption{Minimum model size and robust estimate of standard
deviations (in parentheses).}\label{tb-mms}
\begin{center}
\begin{tabular}{llll}
  \hline
 Model&   NIS & PenGAM & SIS \\
  \hline
Ex 1 ($s=3, SNR\approx 1.01$)&3(0)&3(0)&3(0)\\
Ex 1 ($s=6, SNR\approx1.99$ )&56(0)&1000(0)&56(0)\\
Ex 1 ($s=12, SNR\approx4.07$)&66(7)&1000(0)&62(1)\\
Ex 1 ($s=24, SNR\approx8.20$)&269(134)&1000(0)&109(43)\\
Ex 2 ($SNR\approx0.83$)&3(0)&3(0)&360(361)\\
   \hline
\end{tabular}
\end{center}
\end{table}

The minimum model size(MMS) for each method and its associated
robust estimate of the standard deviation($RSD=IQR/1.34$) are shown
in Table \ref{tb-mms}. The column ``NIS", ``penGAM", and ``SIS"
summarizes the results on the MMS based on 100 simulations,
respectively for the nonparametric independence screening in the
paper, penalized method for additive model of
\cite{MeierGeerBuhlmann09}, and the linear correlation ranking
method of \cite{FanLv08}. For Example 1, when the nonsparsity size
$s>5$, the irrepresentable condition required for the model
selection consistency of LASSO fails.  For these cases, penGAM fails even to
include the true model until the last step.  In contrast, the
proposed nonparametric independence screening performs reasonably
well. It is also worth noting that SIS performs better than NIS in
the first example, particularly for $s=24$.  This is due to the fact
that the true model is linear and the covariates are jointly
normally distributed, which implies that the marginal projection is
also linear.  In this case, NIS selects variables from $pd_n$
parameters whereas SIS selects only from $p$ parameters.  However,
for the nonlinear problem like Example 2, both
nonlinear method NIS and penGAM behave nicely, whereas SIS fails
badly even though the underlying true model is indeed linear.

\subsection{Comparison of Model Selection and Estimation}\label{sec::simu-INIS}
As in the previous section, we set $n=400$ and $p=1000$ for all the examples to demonstrate the power
of our newly proposed methods INIS and g-INIS.  Here in the NIS step, we fix $d_n=5$ as in the last subsection.
 The number of simulations is 100.  Here, we use five-fold cross
validation in Step 2 of the INIS algorithm. For simplicity of notations, we let
$$g_1(x)=x, \quad g_2(x)=(2x-1)^2, \quad g_3(x)=\frac{\sin(2\pi x)}{2-\sin(2\pi x)}$$
  and $$g_4(x)=0.1\sin(2 \pi x)+0.2\cos(2 \pi x)+0.3\sin(2\pi x)^2+0.4\cos(2\pi x)^3+0.5\sin(2 \pi x)^3.$$

{\bf Example 3}.  Following \cite{MeierGeerBuhlmann09}, we generate the data from the following additive model:
    $$Y=5g_1(X_1)+3g_2(X_2)+4g_3(X_3)+6g_4(X_4)+\sqrt{1.74}\varepsilon$$
The covariates $X=(X_1,\cdots, X_p)^T$ are simulated according to the random effect model
  $$X_j=\frac{W_j+t U}{1+t}, j=1,\cdots, p,$$
where $W_1, \cdots, W_p$ and $U$ are i.i.d. $\mbox{Unif}(0,1)$ and $\varepsilon\sim N(0,1)$. When $t=0$, the covariates are all independent, and when $t=1$ the pairwise correlation of covariates is 0.5.

{\bf Example 4}.  Again, we adapt the simulation model from
\cite{MeierGeerBuhlmann09}. This example is a more difficult case than Example 3 since it has 12 important variables with different coefficients.
  \begin{eqnarray*}
    Y&=&g_1(X_1)+g_2(X_2)+g_3(X_3)+g_4(X_4)\\
&+&1.5g_1(X_5)+1.5g_2(X_6)+1.5g_3(X_7)+1.5g_4(X_8)\\
&+&2g_1(X_9)+2g_2(X_{10})+2g_3(X_{11})+2g_4(X_{12})+\sqrt{0.5184}\varepsilon,
  \end{eqnarray*}
where  $\varepsilon\sim N(0,1)$.   The covariates are simulated as in Example 3.

{\bf Example 5}.  We follow the simulation model of \cite{Fan08}, in which
$Y=\beta_1X_1+\beta_2X_2+\beta_3X_3+\beta_4X_4+\varepsilon$ is simulated, where $\varepsilon\sim N(0,1)$. The covariates
$X_1, \cdots, X_p$ are jointly Gaussian, marginally $N(0,1)$, and with $\mbox{corr}(X_i,X_4)=1/\sqrt{2}$ for all $i\neq 4$ and $\mbox{corr}(X_i,X_j)=1/2$ if $i$ and $j$ are distinct elements of $\{1,\cdots, p\} \backslash \{4\}$.  The coefficients
$\beta_1=2, \beta_2=2, \beta_3=2, \beta_4=-3\sqrt{2}$, and
$\beta_j=0$ for $j>4$ are taken so that $X_4$ is independent of $Y$, even though it is the most important variable in the joint model, in terms of the regression coefficient.

For each example, we compare the performances of INIS-penGAM, g-INIS-penGAM
proposed in the paper, penGAM\citep{MeierGeerBuhlmann09},
and ISIS-SCAD \citep{Fan08} which aims for sparse linear model. Their
results are shown respectively in the rows ``INIS'', ``g-INIS'', ``penGAM'' and ``ISIS'' of Table 2, in which the True Positives(TP), False
Positives(FP), Prediction Error(PE) and Computation Time (Time) are reported for each
method. Here the prediction error is calculated on an independent
test data set of size $n/2$.

\begin{table}\label{tb-TPFPPE}
\caption{Average values of the numbers of true (TP) and false (FP)
positives, prediction error (PE), and Time (in seconds). Robust standard deviations are given in parentheses.}
\begin{center}
\begin{tabular}{llllll}
  \hline
 Model& Method & TP&FP&PE&Time\\
\hline&INIS&4.00(0.00)&2.58(2.24)&3.02(0.34)&18.50(7.22)\\
Ex 3($t=0$)&g-INIS&4.00(0.00)&0.67(0.75)&2.92(0.30)&25.03(4.87)\\
($SNR\approx9.02$)&penGAM&4.00(0.00)&31.86(23.51)&3.30(0.40)&180.63(6.92)\\
&ISIS&3.03(0.00)&29.97(0.00)&15.95(1.74)&12.95(4.18)\\
\hline&INIS&3.98(0.00)&15.76(6.72)&2.97(0.39)&78.80(26.91)\\
Ex 3($t=1$)&g-INIS&4.00(0.00)&0.98(1.49)&2.61(0.26)&33.89(9.99)\\
($SNR\approx7.58$)&penGAM&4.00(0.00)&39.21(24.63)&2.97(0.28)&254.06(13.06)\\
&ISIS&3.01(0.00)&29.99(0.00)&12.91(1.39)&18.59(4.37)\\
\hline&INIS&11.97(0.00)&3.22(1.49)&0.97(0.11)&73.60(25.77)\\
Ex 4($t=0$)&g-INIS&12.00(0.00)&0.73(0.75)&0.91(0.10)&160.75(19.94)\\
($SNR\approx8.67$)&penGAM&11.99(0.00)&80.10(18.28)&1.27(0.14)&233.72(10.25)\\
&ISIS&7.96(0.75)&25.04(0.75)&4.70(0.40)&12.89(5.00)\\
\hline&INIS&10.01(1.49)&15.56(0.93)&1.03(0.13)&125.11(39.99)\\
Ex 4($t=1$)&g-INIS&10.78(0.75)&1.08(1.49)&0.87(0.11)&156.37(28.58)\\
($SNR\approx10.89$)&penGAM&10.51(0.75)&62.11(26.31)&1.13(0.12)&278.61(16.93)\\
&ISIS&6.53(0.75)&26.47(0.75)&4.30(0.44)&17.02(4.01)\\
\hline&INIS&3.99(0.00)&21.96(0.00)&1.62(0.18)&94.50(7.12)\\
Ex 5&g-INIS&4.00(0.00)&1.04(1.49)&1.16(0.12)&39.78(12.45)\\
($SNR\approx6.11$)&penGAM&3.00(0.00)&195.03(21.08)&1.93(0.28)&1481.12(181.93)\\
&ISIS&4.00(0.00)&29.00(0.00)&1.40(0.17)&17.78(3.85)\\
\hline
\end{tabular}
\end{center}
\end{table}
First of all, for the greedy modification, g-INIS-penGAM, the number of false positive variables is approximately 1 for all examples and
the number of false positive for both INIS-penGAM and ISIS-SCAD are
much smaller than that for penGAM.  In terms of false positives, we
can see that in Examples 3 and 4, INIS-penGAM and penGAM have
similar performance, whereas penGAM misses one
variable most of the time in Example 5. The linear method ISIS-SCAD missed important
variables in the nonlinear models in Examples 3 and 4.

One may notice that in Example 4 ($t=1$), even INIS and g-INIS miss more than one variables on average. To explore the reason, we took a close look at the iterative process for this example and find out the variable $X_1$ and $X_2$ are missed quite often. The explanation is that although the overall SNR (Signal to Noise Ratio) for this example is around 10.89, the individual contributions to the total signal vary significantly. Now, let us introduce the notion of individual SNR. For example, $\var(m_1(X_1))/\var(\varepsilon)$ in the additive model
$$
   Y = m_1(X_1) + \cdots + m_p(X_p) + \varepsilon
$$
is the individual SNR for the first component under the oracle model where $m_2, \cdots, m_p$ are known. In Example 4 ($t=1$), the variance of all 12 components are as follows:
\begin{table}[ht]
\begin{center}
\begin{tabular}{rrrrrrrrrrrr}
\hline
1&2&3&4&5&6&7&8&9&10&11&12\\
0.08&0.09&0.21&0.26&0.19&0.20&0.47&0.58&0.33&0.36&0.84&1.03\\
\hline
\end{tabular}
\end{center}
\end{table}

We can see that the variance varies a lot among the 12 components, which leads to very different marginal SNRs. For example, the individual SNR for the first component is merely $0.08/0.518 = 0.154$, which is very challenging to be detected. With the overall SNR fixed, the individual SNRs play an important role in measuring the difficulty for selecting individual variables

In the perspective of the prediction error, INIS-penGAM, g-INIS-penGAM and penGAM
outperforms ISIS-SCAD in the nonlinear models whereas their
performances are worse than ISIS-SCAD in the linear model, Example
5.  Overall, it is quite clear that the greedy modification g-INIS is a competitive variable selection method in ultra-high dimensional additive models where we have very low false selection rate, small prediction errors, and fast computation.

\subsection{$d_n$ and SNR }
In this subsection, we conduct simulation study to investigate the performance of INIS-penGAM estimator under different SNR settings using different number ($d_n$) of basis functions.

{\bf Example 6}.  We generate the data from the following additive model:
    $$Y=3g_1(X_1)+3g_2(X_2)+2g_3(X_3)+2g_4(X_4)+ C \sqrt{3.3843}\varepsilon,$$
where the covariates $X=(X_1,\cdots, X_p)^T$ are simulated according to Example 3.  Here $C$ takes a series of different values ($C^2 = 2, 1, 0.5, 0. 25$) to make the corresponding $SNR= 0.5, 1, 2, 4$. We report the results of using number of basis functions $d_n=2, 4, 6, 8$, in Tables 4 and 5 in the Appendix.

From Table \ref{tb-SNR-t=0} in the Appendix where all the variables are independent, both methods have very good true positives under various SNR when $d_n$ is not too large. However, for the case of SNR $=0.5$ and $d_n=16$, the INIS and penGAM perform poorly in terms of low true positive rate. This is due to the fact that when $d_n$ is large, the estimation variance will be large and this makes it difficult to differentiate the active variables from inactive ones when the signals are weak.

Now let us have a look at the more difficult case in Table \ref{tb-SNR-t=1} (in the Appendix) where pairwise correlation between variables is 0.5. We can see that INIS have a competitive performance under various SNR values except when $d_n = 16$.  When SNR $=0.5$, we can not achieve sure screening under the current sample size and configuration for the aforementioned reasons.

\subsection{An analysis on Affymetric
GeneChip Rat Genome 230 2.0 Array}
We use the data set reported in \citet{Scheetz2006} and analyzed by \citet{Huang09} to illustrate
 the application of the proposed method. For this data set, 120 twelve-week-old male rats were selected
for tissue harvesting from the eyes and for microarray analysis. The microarrays used to analyze
the RNA from the eyes of these animals contain over 31,042 different probe sets (Affymetric
GeneChip Rat Genome 230 2.0 Array). The intensity values were normalized using the robust
multi-chip averaging method \citep{Iriz:Hobb:Coll:Beaz:Anto:Sche:Spee:expl:2003} method to obtain summary expression values
for each probe set. Gene expression levels were analyzed on a logarithmic scale.

Following \citet{Huang09}, we are interested in finding the genes that are related to the gene TRIM32, which was
recently found to cause Bardet-Biedl syndrome \citep{Chiang2006}, and is a genetically
heterogeneous disease of multiple organ systems including the retina. Although over 30,000 probe
sets are represented on the Rat Genome 230 2.0 Array, many of them are not expressed in the eye
tissue. We only focus on the 18975 probes which are expressed in the eye tissue. We use our INIS-penGAM method directly on this dataset, where $n=120$ and $p=18975$, and the method is denoted as INIS-penGAM ($p=18975$). Direct application of penGAM approach on the whole dataset is too slow.
Following \citet{Huang09}, we use 2000 probe sets that are expressed in the eye and have highest marginal correlation with TRIM32
in the analysis.  On the subset of the data ($n=120, p=2000$), we apply the INIS-penGAM and penGAM to model the relation between the expression of TRIM32 and those of the
2000 genes.
\begin{figure}[ht]
\caption{Fitted regression functions for the 8 probes that are selected by INIS-penGAM ($p=18975$).}\label{figure:real:rat}

   \centering{\rotatebox{270}{\includegraphics[scale=0.5]{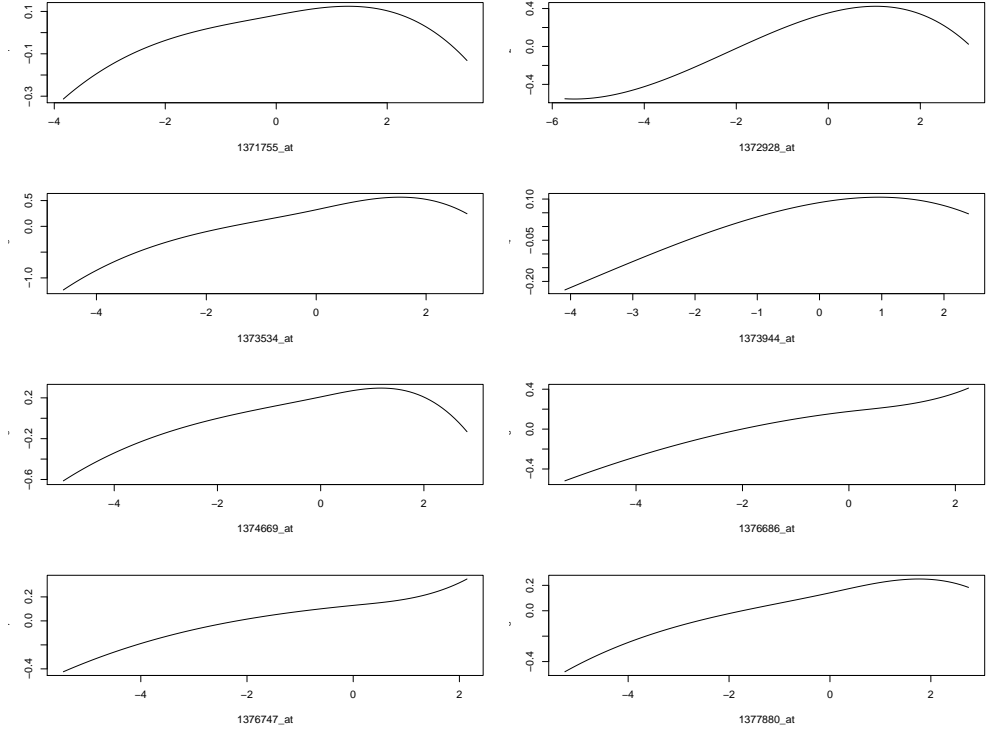}}}

\end{figure}
For simplicity, we did not implement g-INIS-penGAM.   Prior to the analysis, we standardize each probe to be of mean 0 and variance 1. Now, we have three different estimators, INIS-penGAM ($p=18975$), INIS-penGAM ($p=2000$) and penGAM ($p=2000$). The INIS-penGAM ($p=18975$) selects the following 8 probes:
1371755\_at, 1372928\_at,
1373534\_at, 1373944\_at,
1374669\_at, 1376686\_at,
1376747\_at, 1377880\_at.
The INIS-penGAM ($p=2000$) selects the following 8 probes: 1376686\_at,
1376747\_at,
1378590\_at, 1373534\_at,
1377880\_at,
1372928\_at,
1374669\_at,
1373944\_at. On the other hand, the penGAM ($p=2000$) selects 32 probes.
The residual sum of squares (RSS) for these fittings are 0.24, 0.26 and 0.1 for INIS-penGAM ($p=18975$), INIS-penGAM ($p=2000$) and penGAM ($p=2000$), respectively.

\begin{table}[t]
\caption{Mean Model Size (MS) and Prediction Error (PE) over 100 repetitions  and their robust standard deviations(in parentheses) for INIS ($p=18975$), INIS ($p=2000$) and penGAM ($p=2000$).}\label{tb-rat}
\begin{center}
\begin{tabular}{llll}
  \hline
 Method&   MS & PE \\
  \hline
INIS ($p=18975$)&  7.73(0.00) & 0.47(0.13)\\
INIS ($p=2000$)& 7.68(0.75)& 0.44(0.15)\\
penGAM ($p=2000$)& 26.71(14.93)& 0.48(0.16)\\
   \hline
\end{tabular}
\end{center}
\end{table}

In order to further evaluate the performances of the two methods, we use cross-validation and compare the
prediction mean square error (PE). We randomly partition the data into a training set of 100 observations and a test set of 20 observations. We compute the number of probes selected using the 100 observations and the prediction errors on these 20 test sets. This process is repeated 100 times. Table \ref{tb-rat} gives the average values and their associated robust standard deviations over 100 replications. It is clear in the table that by applying the INIS-penGAM approach, we select far fewer genes and give smaller prediction error.  Therefore, in this example, the INIS-penGAM provides the biological investigator a more targeted list of probe sets, which could be very useful in further study.

\section{Discussion}
In this paper, we study the nonparametric independence screening (NIS) method for variable
selection in additive models. B-spline basis functions are used for fitting the marginal
nonparametric components. The proposed marginal projection criteria is an important extension of
the marginal correlation. Iterative NIS procedures are also proposed such that variable selection
and coefficient estimation can be achieved simultaneously. By applying the INIS-penGAM method, we
can preserve the sure screening property and substantially reduce the false selection rate. A greedy modification of the method g-INIS-penGAM is proposed to further reduce the false selection rate.
Moreover, we can deal with the case where some variable is marginally uncorrelated but jointly
correlated with the response. The proposed method can be easily generalized to generalized additive
model with appropriate conditions.

As the additive components are specifically approximated by truncated series expansions with B-spline bases in this paper,
the theoretical results should hold in general and the proposed framework can be readily adaptive to other smoothing methods with additive models
\citep{Horo:Klem:Mamm:opti:2006,Silverman84}, such as local polynomial regression \citep{FanJiang05}, wavelets approximations\citep{AntoFan01,Sardy04} and
smoothing spline \citep{Speckman85}. This is an interesting topic for future research.

\section{Proofs}

{\it Proof of Lemma \ref{lem-0}.}

By the property of the least-squares, $E(Y-f_{nj})f_{nj}=0$ and
$E(Y-f_j)f_{nj}=0$. Therefore, \beqn Ef_{nj}(f_j -f_{nj}) = E(Y-f_{nj})f_{nj} - E(Y-f_j)f_{nj}=0.
\eeqn It follows from this and the orthogonal decomposition $f_j=f_{nj}+(f_j-f_{nj})$ that \beqn
\|f_{nj}\|^2 = \|f_j\|^2  - \|f_j-f_{nj}\|^2. \eeqn The desired result follows from Condition C
together with Fact 1. $\Box$
\bigskip

The following two types of Bernstein's inequality in \citet{vw96} will be needed. We reproduce them
here for the sake of readability.
\begin{lemma}[Bernstein's inequality, Lemma 2.2.9, \citet{vw96}]\label{lem-1}
For independent random variables $Y_1,\cdots, Y_n$ with bounded ranges $[-M,M]$ and zero means,
$$P\left(|Y_1+\cdots+Y_n|>x\right)\leq 2\exp\{-x^2/(2(v+Mx/3))
\},$$
for $v\geq \mbox{var}(Y_1+\cdots+Y_n)$.
\end{lemma}
\bigskip
\begin{lemma}[Bernstein's inequality, Lemma 2.2.11, \citet{vw96}]\label{lem-2}
Let $Y_1,\cdots,Y_n$ be independent random variables with zero mean such that $E|Y_i|^m\leq
m!M^{m-2}v_i/2$, for every $m\geq 2$ (and all $i$) and some constants $M$ and $v_i$. Then
$$P\left(|Y_1+\cdots+Y_n|>x\right)\leq 2\exp\{-x^2/(2(v+Mx))\},$$
for $v\geq v_1+\cdots v_n$.
\end{lemma}
\bigskip

The following two lemmas will be needed to prove Theorem
\ref{the-1}.

\begin{lemma}\label{lem-3}
Under Conditions A, B and D, for any $\delta>0$, there exist some
positive constants $c_6$ and $c_7$ such that \beqn
P(|(\bbP_n-E)\Psi_{jk}Y| \ge \delta n^{-1}) \leq
4\exp(-\delta^2/2(c_6 nd_n^{-1} +c_7 \delta)), \eeqn for
$k=1,\cdots,d_n$, $j=1,\cdots,p$.
\end{lemma}

\bigskip
{\it Proof of Lemma \ref{lem-3}.}

Denote by $T_{jki}=
\Psi_{jk}(X_{ij})Y_i - E\Psi_{jk}(X_{ij})Y_i$.
Since $Y_i=m(\bX_i)+\varepsilon_i$,  we can write
$T_{jki}=T_{jki1}+T_{jki2}$, where
\beqn
 T_{jki1}=\Psi_{jk}(X_{ij})m(\bX_i)-E\Psi_{jk}(X_{ij})m(\bX_i),
\eeqn and $T_{jki2}= \Psi_{jk}(X_{ij})\varepsilon_i. $

By Conditions A, B, D and Fact 2, recalling $\| \Psi_{jk} \|_\infty \leq 1$, we have \beq
\label{eq12} |T_{jki1}|\leq 2B_1, \quad \var(T_{jki1})\le E\Psi_{jk}^2(X_{ij})m_i(X_{ij})^2 \leq B_1^2 C_2
d_n^{-1}. \eeq By Bernstein's inequality (Lemma \ref{lem-1}), for any $\delta_1>0$,
\beq\label{eq13}
  P(\Bigl|\sum_{i=1}^nT_{jki1}\Bigr|>\delta_1)\leq
  2\exp\Bigl(-\frac{1}{2}\frac{\delta_1^2}{nB_1^2C_2d_n^{-1}+2B_1\delta_1/3}\Bigr).
\eeq

Next, we bound the tails of $T_{jki2}$. For every $r \ge 2$,
\beqn
E|T_{jki2}|^r
    & \leq & E|\Psi_{jk}(X_{ij})|^2 E (|\varepsilon_i|^r|\bX_i) \\
    & \leq & r! B_2^{-r} E|\Psi_{jk}(X_{ij})|^2 E \exp( B_2 |\varepsilon_i| | \bX_i) \\
    & \leq & B_3 C_2 d_n^{-1} r! B_2^{-r},
\eeqn where the last inequality utilizes Condition E and Fact 2. By Bernstein's inequality (Lemma
\ref{lem-2}), for any $\delta_2>0$, \beq\label{eq14}
  P(\Bigl| \sum_{i=1}^nT_{jki2} \Bigr|>\delta_2)\leq
  2\exp \Bigl(-\frac{1}{2}\frac{\delta_2^2}{2nB_2^{-2}B_3C_2d_n^{-1}+ B_2^{-1}\delta_2}\Bigr).
\eeq
Combining (\ref{eq13}) and (\ref{eq14}), the desired result follows by taking $c_6=\max(B_1^2C_2,
2B_2^{-2}B_3C_2)$ and $c_7=\max(2/3B_1, B_2^{-1})$. $\Box$
\bigskip

Throughout the rest of the proof, for any matrix $\bA$, let
$\|\bA\|=\sqrt{\lambda_{\max}(\bA^T\bA)}$ be the operator norm and
$\|\bA\|_{\infty}=\max_{i, j} |A_{ij}|$ be the infinity norm. The next lemma
is about the tail probability of the eigenvalues of the design
matrix.

\begin{lemma}\label{lem-4}
Under Conditions A and B, for any $\delta>0$,
\beqn
&&P(|\lambda_{\min}(\bbP_n\bPsi_j\bPsi_j^T) - \lambda_{\min}(E\bPsi_j\bPsi_j^T)|
\ge d_n \delta/n) \\
&\le& 2 d_n^2
\exp\Bigl\{-\frac{1}{2}\frac{\delta^2}{C_2 nd_n^{-1} + \delta/3}\Bigr\}.
\eeqn In addition, for any given constant $c_4$, there exists some
positive constant $c_{8}$ such that \beq\label{eq15}
&&P\left\{\Bigl|\left\|(\mathbb{P}_n \bPsi_j \bPsi_j^T)^{-1}\right\|
- \left\|(E \bPsi_j \bPsi_j^T)^{-1}\right\|\Bigr| \ge
c_{8}\left\|(E\bPsi_j\bPsi_j^T)^{-1}\right\|\right\} \nonumber\\
&\le& 2
d_n^2\exp\Bigl(-c_4nd_n^{-3} \Bigr).
\eeq
\end{lemma}

{\it Proof of Lemma \ref{lem-4}}.

For any symmetric matrices $\bA$ and $\bB$ and any $\|\bx\|=1$, where $\|\cdot\|$ is the Euclidean norm,
$$\bx^T(\bA+\bB)x=\bx^T\bA \bx+\bx^T\bB \bx\geq \min_{\|\bx\|=1}\bx^T\bA \bx+\min_{\|\bx\|=1}\bx^T\bB
\bx.$$
Taking minimum among $\|\bx\|=1$ on the left side, we have
$$\min_{\|\bx\|=1}\bx^T(\bA+\bB)\bx\geq \min_{\|\bx\|=1}\bx^T\bA \bx+\min_{\|\bx\|=1}\bx^T\bB \bx, $$
which is equivalent to $\lambda_{\min}(\bA+\bB)\geq
\lambda_{\min}(\bA)+\lambda_{\min}(\bB)$.

Then we have
$$\lambda_{\min}(\bA)\geq\lambda_{\min}(\bB)+\lambda_{\min}(\bA-\bB),$$
which is the same as $$\lambda_{\min}(\bA-\bB)\leq
\lambda_{\min}(\bA)-\lambda_{\min}(\bB).$$

By switching the roles of $\bA$ and $\bB$, we also have
$$
       \lambda_{\min} (\bB - \bA) \leq \lambda_{\min} (\bB)  - \lambda_{\min}(\bA)
$$
In other words,
\begin{eqnarray}\label{eq16}
  | \lambda_{\min} (\bA)  - \lambda_{\min}(\bB)| \leq \max\{|\lambda_{\min} (\bA - \bB)|, |\lambda_{\min} (\bB - \bA)|\}
\end{eqnarray}

Let $\bD_j = \bbP_n\bPsi_j\bPsi_j^T-E\bPsi_j\bPsi_j^T$.  Then, it
follows from (\ref{eq16}) that \beq\label{eq17}
|\lambda_{\min}(\bbP_n\bPsi_j\bPsi_j^T) -
\lambda_{\min}(E\bPsi_j\bPsi_j^T) | \leq   \max\{|\lambda_{\min}
(\bD_j)|, |\lambda_{\min} (-\bD_j)|\}. \eeq We now bound the
right-hand side of (\ref{eq17}).  Let $\bD_j^{(i,l)}$ be the $(i,l)$
entry of $\bD_j$.  Then, it is easy to see that for any $\|\bx\|=
1$, \beq \label{eq18} |\bx^T \bD_j \bx| \leq \|\bD_j\|_{\infty}
\Bigl(\sum_{i=1}^{d_n}|x_i| \Bigr)^2 \leq d_n \|\bD_j\|_{\infty}.
\eeq

Thus,
\beqn
\lambda_{\min}(\bD_j) = \min_{\|\bx \|=1} \bx^T \bD_j \bx
  \leq d_n \|\bD_j\|_{\infty}.
\eeqn On the other hand, by using (\ref{eq18}) again, we have
$$
  \lambda_{\min}(\bD_j) = - \max_{\|\bx \|=1} (- \bx^T \bD_j \bx)
  \geq - d_n \|\bD_j\|_{\infty}.
$$
We conclude that
$$
    | \lambda_{\min}(\bD_j) | \leq  d_n \|\bD_j\|_{\infty}.
$$
The same bound on $|\lambda_{\min}(-\bD_j)|$ can be obtained by
using the same argument.  Thus, by (\ref{eq17}), we have \beq
\label{eq19} |\lambda_{\min}(\bbP_n\bPsi_j\bPsi_j^T) -
\lambda_{\min}(E\bPsi_j\bPsi_j^T) | \leq d_n \|\bD_j\|_{\infty}.
\eeq

We now use Bernstein's inequality to bound the right-hand side of (\ref{eq19}). Since
$\|\Psi_{jk}\|_{\infty} \le 1$, and by using Fact 2, we have that \beqn
\var(\Psi_{jk}(X_j)\Psi_{jl}(X_j)) \leq E\Psi_{jk}^2(X_j)\Psi_{jl}^2(X_j)\leq E\Psi_{jk}^2(X_j)\leq
C_2 d_n^{-1}. \eeqn By Bernstein's inequality (Lemma \ref{lem-1}), for any $\delta>0$, \beq
\label{eq20} P(|(\mathbb{P}_n-E) \Psi_{jk}(X_j)\Psi_{jl}(X_j)|> \delta/n) \le 2 \exp
\Bigl\{-\frac{\delta^2} {2(C_2 nd_n^{-1} + \delta/3)} \Bigr\}. \eeq It follows from (\ref{eq19}),
(\ref{eq20}) and the union bound of probability that \beqn
&&P(|\lambda_{\min}(\mathbb{P}_n \bPsi_j \bPsi_j^T) - \lambda_{\min}(E \bPsi_j \bPsi_j^T)| \ge d_n\delta/n)\nonumber\\
& \le& 2d_n^{2} \exp \Bigl\{-\frac{\delta^2}
{2(C_2nd_n^{-1} + \delta/3)} \Bigr\}.
\eeqn
This completes the proof of the first inequality.

To prove the second inequality, let us take $\delta=c_{9}
D_1nd_n^{-2}$ in (\ref{eq20}), where $c_{9} \in (0,1)$.  By recalling Fact 3, it follows that
\beq\label{eq21}
&&P(|\lambda_{\min}(\mathbb{P}_n \bPsi_j \bPsi_j^T) -
\lambda_{\min}(E \bPsi_j \bPsi_j^T)| \ge
 c_{9}\lambda_{\min}(E\bPsi_j\bPsi_j^T)) \nonumber\\
 &\le& 2 d_n^2\exp\Bigl(-c_4 nd_n^{-3}  \Bigr),
\eeq for some positive constant $c_4$.  The second part of the lemma thus follows from the fact that $\lambda_{\min}(\bH)^{-1} = \lambda_{\max}(\bH^{-1})$, if we establish
\beq \label{eq22}
&&P\left(\left|\Bigl\{\lambda_{\min}(\mathbb{P}_n \bPsi_j
\bPsi_j^T)\Bigr\}^{-1} - \Bigl\{\lambda_{\min}(E \bPsi_j \bPsi_j^T)\Bigr\}^{-1}\right| \ge
c_{8 }\Bigl\{\lambda_{\min}(E\bPsi_j\bPsi_j^T)\Bigr\}^{-1}\right)
\nonumber \\
&\le& 2 d_n^2\exp\Bigl(-c_4nd_n^{-3}\Bigr), \eeq by using
(\ref{eq21}), where $c_{8 } = 1/(1-c_{9}) -1 $.

We now deduce (\ref{eq22}) from (\ref{eq21}). Let $A =
\lambda_{\min}(\bbP_n\bPsi_j \bPsi_j^T)$ and $B =
\lambda_{\min}(E\bPsi_j \bPsi_j^T)$. Then, $A>0$ and $B>0$. We aim
to show for $a \in (0,1)$,
$$|A^{-1}-B^{-1}| \ge cB^{-1} ~\mbox{implies} ~|A-B| \ge aB,
$$
where  $c = 1/(1-a) -1$.

Since $$ |A^{-1} - B^{-1}| \ge (1/(1-a) -1)B^{-1},$$ we
have \beqn
 A^{-1} - B^{-1} \le -(1/(1-a) -1) B^{-1},
\quad \mbox{~or~} \quad \ge (1/(1-a) - 1)B^{-1}.\nonumber
\eeqn Note that for $a \in (0, 1)$, we have $1 - 1/(1+a) <
1/(1-a) - 1$.
Then it follows that \beqn
 A^{-1} - B^{-1} \le -(1-1/(1+a)) B^{-1},
\quad \mbox{~or~} \quad \ge (1/(1-a) - 1)B^{-1},\nonumber
\eeqn which is equivalent to $|A-B| \ge aB.$

This concludes the proof of the lemma. $\Box$

%
%

{\it Proof of Theorem \ref{the-1}.}

We first show part (i).
Recall that
\beqn
\|\hat f_{nj} \|_n^2 &=& (\mathbb{P}_n \bPsi_j Y)^T(\mathbb{P}_n \bPsi_j \bPsi_j^T)^{-1}\mathbb{P}_n \bPsi_j Y,
\eeqn
and
\beqn
\|f_{nj} \|^2
&=& (E \bPsi_j Y)^T(E \bPsi_j \bPsi_j^T)^{-1}E \bPsi_j Y.
\eeqn

Let $\ba_n = \bbP_n \bPsi_jY$,
$\bB_n = (\bbP_n \bPsi_j\bPsi_j^T)^{-1}$, $\ba = E \bPsi_jY$ and $\bB=
(E \bPsi_j\bPsi_j^T)^{-1}$. By some algebra,
\beqn
&&\ba_n^T\bB_n\ba_n -
\ba^T\bB\ba = (\ba_n-\ba)^T\bB_n(\ba_n-\ba) + 2(\ba_n-\ba)^T\bB_n\ba +
\ba_n^T(\bB_n-\bB)\ba, \eeqn
we have \beq\label{eq23} \|\hat f_{nj}
\|_n^2 - \| f_{nj}\|^2 = S_1 + S_2 + S_3, \eeq where \beqn
&&S_1 = \Bigl(\A-\ea\Bigr)^T\B\Bigl(\A-\ea\Bigr), \\
&&S_2 = 2\Bigl(\A-\ea\Bigr)^T\B \ea, \\
&&S_3 = (\ea)^T\Bigl(\B-\eb\Bigr)\ea. \eeqn

Note that \beq\label{eq24} S_1 \le  \|\B\| \cdot \|\A - \ea \|^2.
\eeq By Lemma \ref{lem-3} and the union bound of probability,
\beq\label{eq25} P(\|\A - \ea\|^2 \ge d_n\delta^2 n^{-2}) \le
4d_n\exp(-\delta^2/2(c_6 nd_n^{-1} +c_7 \delta)). \eeq Recall the
result in Lemma \ref{lem-4} that, for any given constant $c_4$,
there exists a positive constant $c_8$ such that \beqn
&&P\left\{\Bigl|\|(\mathbb{P}_n \bPsi_j \bPsi_j^T)^{-1}\| - \|(E
\bPsi_j \bPsi_j^T)^{-1}\|\Bigr| \ge
c_{8}\|(E\bPsi_j\bPsi_j^T)^{-1}\|\right\} \nonumber\\
&\le& 2 d_n^2\exp\Bigl(-c_4nd_n^{-3} \Bigr). \eeqn Since by Fact 3,
\beqn \Bigl\|(E\bPsi_j\bPsi_j^T)^{-1}\Bigr\| \le D_1^{-1} d_n, \eeqn
it follows that \beq\label{eq26} P \left \{ \Bigl\|(\mathbb{P}_n \bPsi_j
\bPsi_j^T)^{-1}\Bigr\| \ge (c_{8}+1)D_1^{-1}d_n\right \} \le 2
d_n^2\exp\Bigl(-c_4nd_n^{-3} \Bigr). \eeq

Combining (\ref{eq24})--(\ref{eq26}) and the union bound of
probability, we have \beq\label{eq27} P(S_1 \ge (c_8+1)
D_1^{-1}d_n^2 \delta^2/n^2 ) \le 4d_n\exp(-\delta^2/2(c_6 nd_n^{-1}
+c_7 \delta)) + 2 d_n^2\exp\Bigl(-c_4nd_n^{-3} \Bigr). \eeq

To bound $S_2$, we note that
\beq\label{eq28}
|S_2| &\le& 2\|\A -\ea \|\cdot\|\B\ea \|\nonumber\\
    &\le& 2\|\A -\ea \|\cdot\| \B\|\cdot\|\ea \|.
\eeq
Since by Condition D,
 \beq\label{eq29}
\|E\bPsi_jY \|^2 = \sum_{k=1}^{d_n}(E\Psi_{jk}Y)^2=
\sum_{k=1}^{d_n}(E\Psi_{jk}m)^2 \le
\sum_{k=1}^{d_n}{B_1^2 E\Psi_{jk}^2} \leq B_1^2C_2,
 \eeq
it follows from (\ref{eq25}), (\ref{eq26}), (\ref{eq28}),
(\ref{eq29}) and the union bound of probability that \beq
\label{eq30}
&&P(|S_2| \ge 2(c_8+1){D_1^{-1}C_2^{1/2}}B_1 d_n^{3/2} \delta/n ) \nonumber\\
&& \le 4d_n\exp(-\delta^2/2(c_6 nd_n^{-1} +c_7 \delta))+ 2
d_n^2\exp\Bigl(-c_4nd_n^{-3} \Bigr). \eeq

Now we bound $S_3$. Note that \beq\label{eq31} S_3 = (\ea)^T \B
\Bigl(E-\bbP_n\Bigr) \bPsi_j\bPsi_j^T \eb \ea. \eeq By the fact that
$\|\bA\bB\|\leq\|\bA\|\cdot\|\bB\|$, we have \beq\label{eq32} |S_3|
\le \|(\bbP_n-E) \bPsi_j\bPsi_j^T\|\cdot \|\B\|\cdot\|\eb\|\cdot\|
\ea\|^2. \eeq
For any $\|\bx\|= 1$ and $d_n$-dimensional square
matrix $\bD$, \beqn \bx^T \bD^T \bD \bx = \sum_{i} (\sum_{j} d_{ij} x_j)^2 \leq \|\bD\|_{\infty}^2 d_n
\Bigl(\sum_{j=1}^{d_n}|x_i| \Bigr)^2 \leq d_n^2 \|\bD\|^2_{\infty}.
\eeqn
Therefore, $\|\bD\| \le d_n \|\bD\|_{\infty}$.  We conclude that
\beq\label{eq33}
\left \| \Bigl(\bbP_n-E\Bigr)
\bPsi_j\bPsi_j^T) \right \| \le d_n \|(\bbP_n-E) \bPsi_j\bPsi_j^T\|_{\infty}.
\eeq By (\ref{eq20}), (\ref{eq26}), (\ref{eq29}), (\ref{eq32}),
(\ref{eq33}) and the union bound of probability, it follows that
\beq \label{eq34}
&&P(|S_3| \ge (c_8+1)D_1^{-2}B_1^2C_2 d_n^3 \delta/n ) \nonumber\\
&& \le
2d_n^2\exp(-\delta^2/2(c_6 nd_n^{-1} +c_7 \delta)) +
2d_n^2\exp\Bigl(-c_4nd_n^{-3} \Bigr). \eeq

It follows from (\ref{eq23}), (\ref{eq27}), (\ref{eq30}), (\ref{eq34}) and the union bound of probability that for some positive constants $c_{10}$, $c_{11}$ and $c_{12}$,
\beq\label{eq35}
&&P \left (\Bigl|\| \hat f_{nj}\|_n^2 - \|f_{nj}\|^2\Bigr| \ge
c_{10} d_n^2\delta^2/n^2 + c_{11}d_n^{3/2}\delta/n + c_{12}d_n^3\delta/n \right ) \nonumber\\
&& \le (8d_n + 2d_n^2)\exp(-\delta^2/2(c_6 nd_n^{-1} +c_7 \delta)) +
6d_n^2\exp\Bigl(-c_4nd_n^{-3} \Bigr).
\eeq

In (\ref{eq35}), let $c_{10}d_n^2 \delta^2/n^2 + c_{11}d_n^{3/2}
\delta/n + c_{12}d_n^3 \delta/n = c_2d_nn^{-2\kappa}$ for any given
$c_2>0$, i.e., taking $\delta=n^{1-2\kappa}d_n^{-2}c_2/c_{12}$, there
exist some positive constants $c_3$ and $c_4$ such that \beqn
&& P(\Bigl|\|\hat f_{nj} \|_n^2 - \|f_{nj} \|^2 \Bigr| \ge c_2d_nn^{-2\kappa}) \nonumber\\
&&\le (8d_n+2d_n^2)\exp(-c_3n^{1-4\kappa}d_n^{-3}) + 6d_n^2\exp\Bigl(-c_4nd_n^{-3} \Bigr).
\eeqn
The first part thus follows the union bound of probability.

To prove the second part, note that on the event
$$
    A_n \equiv \{\max_{j \in {\cal M}_\star}  \Bigl|\|\hat{f}_{nj}\|_n^2 -
    \|f_{nj}\|^2 \Bigr| \leq c_1 \xi d_n n^{-2\kappa}/2\},
$$
by Lemma 1, we have
\beq\label{eq36}
\|\hat f_{nj}\|_n^2 \geq c_1 \xi d_n n^{-2\kappa}/2, \quad
      \mbox{for all $j \in {\cal M}_\star$}.
\eeq Hence, by the choice of $\nu_n$, we have ${\cal M}_\star
\subset \widehat {\cal M}_{\nu_n}$.  The result now follows from a
simple union bound: \beqn P(A_n^c) \leq s_n
\Bigl\{(8d_n+2d_n^2)\exp\Bigl(-c_3n^{1-4\kappa}d_n^{-3}\Bigr) +
6d_n^2\exp\Bigl(-c_4nd_n^{-3}\Bigr)\Bigr\}. \eeqn This completes the
proof.  $\Box$

{\it Proof of Theorem \ref{the-2}.}
The key idea of the proof is to show that \beq\label{eq37}
 \| E \bPsi Y \|^2  =
O(\lambda_{\max}(\bSigma)). \eeq If so,  by
definition and $\|\Psi_{jk}\|_\infty \leq 1$,
 we have
 \beqn
\sum_{j=1}^{p_n} \|f_{nj}\|^2 \le \max_{1\le j\le p_n}\lambda_{\max}\{
(E\bPsi_j\bPsi_j^T)^{-1} \}
 \| E \bPsi Y \|^2 =O(d_n\lambda_{\max}(\bSigma)). \eeqn
This implies that the number of $\{j: \|f_{nj} \|^2
> \varepsilon d_n n^{-2\kappa}\}$ can not exceed $O(n^{2\kappa} \lambda_{\max}(\bSigma))$ for any
$\varepsilon>0$. Thus, on the set
$$
  B_n = \{\max_{1 \leq j  \leq p_n} \Bigl|\|\hat f_{nj} \|_n^2- \|f_{nj}\|^2\Bigr| \leq
  \varepsilon d_n
  n^{-2\kappa}\},
$$
the number of $\{j: \|\hat f_{nj}\|_{n}^2 > 2 \varepsilon d_n
n^{-2\kappa}\}$ can not exceed the number of $\{j:
\|f_{nj}\|^2>\varepsilon d_n n^{-2\kappa} \}$, which is bounded by
$O\{ n^{2\kappa} \lambda_{\max} (\bSigma)\}$. By taking $\varepsilon =
c_5/2$, we have
$$
P [ | \widehat {\cal M}_{\nu_n}| \leq O\{ n^{2\kappa} \lambda_{\max}
(\bSigma)\} ] \geq P(B_n).
$$
The conclusion follows from Theorem~\ref{the-1}(i).

It remains to prove (\ref{eq37}). Note that (\ref{eq37}) is more
related to the joint regression rather than the marginal regression.
Let \beqn \balpha_n = \argmin_{\balpha} E \Bigl(Y -
\bPsi^T\balpha\Bigr)^2, \eeqn which is the joint regression
coefficients in the population.
By the score equation of $\balpha_n$, we get \beqn
 E\bPsi(Y - \bPsi^T\balpha_n) = 0.
\eeqn
Hence \beqn \| E \bPsi Y \|^2 =
\balpha_n^{T} E\bPsi\bPsi^T E\bPsi\bPsi^T\balpha_n \leq
\lambda_{\max} (\bSigma) \balpha_n^{T}
E\bPsi\bPsi^T\balpha_n, \eeqn Now,  it follows from the orthogonal
decomposition that
$$
\var(Y) = \var(\bPsi^T\balpha_n) + \var( Y - \bPsi^T\balpha_n).
$$
Since $\var(Y) = O(1)$, we conclude that $\var(\bPsi^T\balpha_n) =
O(1)$, i.e.
$$
\balpha_n^{T} E\bPsi\bPsi^T\balpha_n = O(1).
$$
This completes the proof. $\Box$.

\bigskip

\bibliographystyle{ims}
\bibliography{amsis-ref}

\appendix
\section{APPENDIX: Tables for Simulation Results of Section 5.3}
\begin{table}[!h]
\caption{Average values of the numbers of true (TP), false (FP)
positives, prediction error (PE), computation time (Time) for Example 6 ($t=0$).  Robust standard deviations are given in parentheses.}\label{tb-SNR-t=0}
\begin{center}
\begin{tabular}{lllllll}
  \hline
 SNR&$d_n$& Method & TP&FP&PE &Time\\
\hline
\multirow{8}{*}{0.5}&\multirow{2}{*}{ 2}&INIS& 3.96(0.00)& 2.28(1.49)& 7.74(0.79)& 16.09(5.32)\\
&&penGAM& 4.00(0.00)& 27.85(16.98)& 8.07(0.92)& 354.46(31.48)\\
&\multirow{2}{*}{ 4}&INIS& 3.93(0.00)& 2.29(1.68)& 7.90(0.81)& 21.68(8.95)\\
&&penGAM& 3.99(0.00)& 25.61(13.62)& 8.21(0.84)& 421.17(35.71)\\
&\multirow{2}{*}{ 8}&INIS& 3.81(0.00)& 2.59(2.24)& 8.16(1.08)& 33.10(15.79)\\
&&penGAM& 3.95(0.00)& 34.59(20.34)& 8.49(0.82)& 484.17(179.70)\\
&\multirow{2}{*}{16}&INIS& 3.38(0.75)& 2.02(1.49)& 8.60(1.13)& 42.69(20.13)\\
&&penGAM& 3.74(0.00)& 33.48(23.88)& 9.04(0.93)& 685.97(267.43)\\
\hline
\multirow{8}{*}{1.0}&\multirow{2}{*}{ 2}&INIS& 4.00(0.00)& 2.16(2.24)& 3.98(0.34)& 16.03(5.74)\\
&&penGAM& 4.00(0.00)& 26.51(14.18)& 4.20(0.46)& 284.85(20.30)\\
&\multirow{2}{*}{ 4}&INIS& 4.00(0.00)& 2.08(1.49)& 3.97(0.45)& 20.80(8.57)\\
&&penGAM& 4.00(0.00)& 28.33(15.49)& 4.24(0.47)& 362.02(81.43)\\
&\multirow{2}{*}{ 8}&INIS& 4.00(0.00)& 2.72(2.24)& 4.04(0.43)& 35.79(18.38)\\
&&penGAM& 4.00(0.00)& 36.50(21.83)& 4.37(0.47)& 427.60(152.53)\\
&\multirow{2}{*}{16}&INIS& 4.00(0.00)& 1.80(1.49)& 4.26(0.45)& 46.81(21.47)\\
&&penGAM& 4.00(0.00)& 38.60(19.78)& 4.80(0.57)& 595.87(197.06)\\
\hline
\multirow{8}{*}{2.0}&\multirow{2}{*}{ 2}&INIS& 4.00(0.00)& 2.03(2.24)& 2.12(0.17)& 15.92(5.42)\\
&&penGAM& 4.00(0.00)& 25.89(13.06)& 2.25(0.24)& 235.69(13.32)\\
&\multirow{2}{*}{ 4}&INIS& 4.00(0.00)& 2.38(2.24)& 2.06(0.22)& 23.54(9.08)\\
&&penGAM& 4.00(0.00)& 30.37(17.16)& 2.21(0.26)& 341.13(19.44)\\
&\multirow{2}{*}{ 8}&INIS& 4.00(0.00)& 2.79(2.24)& 2.03(0.21)& 38.56(19.58)\\
&&penGAM& 4.00(0.00)& 38.51(16.42)& 2.24(0.26)& 396.84(20.51)\\
&\multirow{2}{*}{16}&INIS& 4.00(0.00)& 1.77(1.49)& 2.17(0.25)& 48.40(24.65)\\
&&penGAM& 4.00(0.00)& 42.58(16.60)& 2.54(0.30)& 540.89(165.39)\\
\hline
\multirow{8}{*}{4.0}&\multirow{2}{*}{ 2}&INIS& 4.00(0.00)& 2.06(2.24)& 1.19(0.13)& 17.74(6.42)\\
&&penGAM& 4.00(0.00)& 28.57(14.37)& 1.27(0.15)& 213.43(12.09)\\
&\multirow{2}{*}{ 4}&INIS& 4.00(0.00)& 2.33(1.49)& 1.09(0.10)& 23.28(9.37)\\
&&penGAM& 4.00(0.00)& 30.75(17.35)& 1.18(0.14)& 300.69(12.21)\\
&\multirow{2}{*}{ 8}&INIS& 4.00(0.00)& 2.88(2.24)& 1.02(0.12)& 39.21(19.17)\\
&&penGAM& 4.00(0.00)& 40.51(17.54)& 1.14(0.14)& 340.06(11.49)\\
&\multirow{2}{*}{16}&INIS& 4.00(0.00)& 1.72(1.49)& 1.10(0.12)& 49.79(25.78)\\
&&penGAM& 4.00(0.00)& 45.77(19.03)& 1.33(0.16)& 481.19(141.51)\\
\hline
\end{tabular}
\end{center}
\end{table}

\begin{table}[!h]
\caption{Average values of the numbers of true (TP), false (FP)
positives, prediction error (PE), computation time (Time) for Example 6 ($t=1$).  Robust standard deviations are given in parentheses. }\label{tb-SNR-t=1}
\begin{center}
\begin{tabular}{lllllll}
  \hline
 SNR&$d_n$& Method & TP&FP&PE &Time\\
\hline
\multirow{8}{*}{0.5}&\multirow{2}{*}{ 2}&INIS& 3.35(0.75)& 33.67(8.96)& 9.49(1.28)& 196.87(91.48)\\
&&penGAM& 3.10(0.00)& 17.74(15.11)& 7.92(0.89)& 1107.78(385.95)\\
&\multirow{2}{*}{ 4}&INIS& 3.02(0.00)& 20.22(2.43)& 8.70(1.14)& 109.51(56.11)\\
&&penGAM& 2.78(0.00)& 15.91(10.07)& 7.99(0.91)& 734.08(227.55)\\
&\multirow{2}{*}{ 8}&INIS& 2.51(0.75)& 10.48(0.75)& 8.37(0.89)& 65.12(16.64)\\
&&penGAM& 2.59(0.75)& 16.47(9.70)& 8.13(0.90)& 624.31(56.23)\\
&\multirow{2}{*}{16}&INIS& 2.10(0.00)& 4.47(0.75)& 8.44(1.00)& 46.84(15.61)\\
&&penGAM& 2.41(0.75)& 15.56(10.63)& 8.42(0.97)& 786.45(244.02)\\
\hline
\multirow{8}{*}{1.0}&\multirow{2}{*}{ 2}&INIS& 3.83(0.00)& 32.46(9.70)& 4.86(0.60)& 164.97(64.14)\\
&&penGAM& 3.64(0.75)& 24.61(21.08)& 4.19(0.49)& 849.23(294.03)\\
&\multirow{2}{*}{ 4}&INIS& 3.56(0.75)& 20.53(1.68)& 4.42(0.52)& 118.14(43.97)\\
&&penGAM& 3.46(0.75)& 22.07(16.04)& 4.18(0.49)& 614.93(97.36)\\
&\multirow{2}{*}{ 8}&INIS& 3.09(0.00)& 10.67(0.75)& 4.28(0.49)& 71.16(32.10)\\
&&penGAM& 3.12(0.00)& 19.92(10.63)& 4.30(0.50)& 548.60(33.88)\\
&\multirow{2}{*}{16}&INIS& 2.68(0.75)& 4.18(0.75)& 4.45(0.52)& 46.08(15.35)\\
&&penGAM& 2.95(0.00)& 16.39(11.19)& 4.57(0.55)& 710.56(199.86)\\
\hline
\multirow{8}{*}{2.0}&\multirow{2}{*}{ 2}&INIS& 3.99(0.00)& 29.45(11.57)& 2.55(0.38)& 139.67(70.45)\\
&&penGAM& 3.97(0.00)& 36.57(22.57)& 2.25(0.28)& 626.84(210.44)\\
&\multirow{2}{*}{ 4}&INIS& 3.93(0.00)& 19.12(3.73)& 2.26(0.24)& 111.01(21.82)\\
&&penGAM& 3.91(0.00)& 31.31(20.52)& 2.19(0.23)& 481.87(52.11)\\
&\multirow{2}{*}{ 8}&INIS& 3.50(0.75)& 10.29(0.75)& 2.21(0.23)& 78.06(32.23)\\
&&penGAM& 3.71(0.75)& 27.06(19.03)& 2.28(0.29)& 448.38(26.63)\\
&\multirow{2}{*}{16}&INIS& 2.93(0.00)& 4.07(0.00)& 2.42(0.32)& 51.69(1.10)\\
&&penGAM& 3.22(0.00)& 19.51(12.13)& 2.53(0.30)& 661.93(46.27)\\
\hline
\multirow{8}{*}{4.0}&\multirow{2}{*}{ 2}&INIS& 4.00(0.00)& 29.47(11.38)& 1.45(0.21)& 144.22(72.54)\\
&&penGAM& 4.00(0.00)& 37.27(20.71)& 1.27(0.17)& 533.98(69.29)\\
&\multirow{2}{*}{ 4}&INIS& 3.99(0.00)& 17.36(5.22)& 1.17(0.12)& 102.97(32.71)\\
&&penGAM& 4.00(0.00)& 38.71(20.34)& 1.16(0.11)& 403.32(28.29)\\
&\multirow{2}{*}{ 8}&INIS& 3.78(0.00)& 10.00(0.00)& 1.13(0.16)& 88.79(12.02)\\
&&penGAM& 3.99(0.00)& 41.42(15.86)& 1.19(0.13)& 402.92(16.94)\\
&\multirow{2}{*}{16}&INIS& 3.02(0.00)& 3.98(0.00)& 1.36(0.15)& 49.13(1.85)\\
&&penGAM& 3.72(0.75)& 29.58(19.40)& 1.43(0.18)& 556.31(35.48)\\
\hline
\end{tabular}
\end{center}
\end{table}
\end{document}